\documentclass[12pt]{article}
\usepackage[english]{babel}
\usepackage{graphicx,epsfig}
\makeindex \textwidth 170mm \textheight 220mm \topmargin -5mm
\newcommand{\be}{\begin{equation}}
\newcommand{\ee}{\end{equation}}
\newcommand{\bea}{\begin{eqnarray}}
\newcommand{\eea}{\end{eqnarray}}

\def\sus#1{\sin^{#1}i}
\def\jw2{w_{.2}}
\def\ecn#1{(1-e^{2})^{#1}}
\def\rtr#1{\left(\frac{R}{a}\right)^{#1}}
\def\jn2{\dot\Omega_{.2}}
\def\jo2{\dot\omega_{.2}}
\def\ekn#1#2{(1+\frac{#1}{#2}e^{2})}
\def\djn{\dot\Omega_{.2n}\equiv\derp{{\dot\Omega^{\rm even\ zonal}_{\rm class}}}{(J_{2n})}}
\def\djo{\dot\omega_{.2n}\equiv\derp{{\dot\omega^{\rm even\ zonal}_{\rm class}}}{(J_{2n})}}
\def\jn#1{\dot\Omega_{.#1}}
\def\jo#1{\dot\omega_{.#1}}
\def\eqiaz{\begin{eqnarray*}}
\def\eqfaz{\end{eqnarray*}}
\def\eqia{\begin{eqnarray}}
\def\eqfa{\end{eqnarray}}
\def\btab{\begin{tabular}}
\def\etab{\end{tabular}}
\def\bar{\begin{array}}
\def\ear{\end{array}}

\def\btab{\begin{tabular}}
\def\etab{\end{tabular}}
\def\rfr#1{Equation (\ref{#1})}

\def\bb{\bibitem{}}
\def\bar{\begin{array}}
\def\ear{\end{array}}
\def\eqi{\begin{equation}}
\def\eqf{\end{equation}}

\def\og{\omega}

\def\O{\Omega}

\def\ci{\cos{i}}

\def\vass#1{\left\vert\ #1 \right\vert}
\def\rp#1#2{{#1\over#2}}

\def\derp#1#2{\rp{\partial{#1}}{\partial{#2}}}
\def\dert#1#2{\frac{{{d}}{#1}}{{{d}}{#2}}}

\def\lb#1{\label{#1}}

\def\derp#1#2{\rp{\partial{#1}}{\partial{#2}}}

\def\lb#1{\label{#1}}

\def\ci{\cos{i}}

\def\ci{\cos{i}}

\def\dman2{-\sqrt{1-e^2}\ (\dert{\og}{t}+\ci\dert{\O}{t})}

\def\btab{\begin{tabular}}
\def\etab{\end{tabular}}

\def\btab{\begin{tabular}}
\def\etab{\end{tabular}}
\def\bar{\begin{array}}
\def\ear{\end{array}}
\def\dert#1#2{\rp{{\rm d}{#1}}{{\rm d}{#2}}}

\def\grl{general relativistic}

\def\leti{\anx{Lense-Thirring}}

\def\lb#1{\label{#1}}

\def\rp#1#2{{#1\over#2}}

\def\rfr#1{eq. (\ref{#1})}

\def\bb{\bibitem}
\def\eqi{\begin{equation}}
\def\eqf{\end{equation}}
\def\eqia{\begin{eqnarray}}
\def\eqfa{\end{eqnarray}}
\def\btab{\begin{tabular}}
\def\etab{\end{tabular}}
\def\bar{\begin{array}}
\def\ear{\end{array}}
\def\dert#1#2{\rp{{d}{#1}}{{d}{#2}}}

\def\grl{general relativistic}

\def\leti{Lense--Thirring}

\def\lb#1{\label{#1}}

\def\rp#1#2{{#1\over#2}}

\begin{document}
\begin{titlepage}
\begin{flushright}
\today\\
BARI-TH/00\\
\end{flushright}
\vspace{.5cm}
\begin{center}
{\LARGE The impact of the static part of the Earth's gravity field
on some tests of General Relativity with Satellite Laser Ranging}
\vspace{1.0cm}
\quad\\
{Lorenzo Iorio$^{\dag}$\\ \vspace{0.5cm}
\quad\\
{\dag}Dipartimento di Fisica dell' Universit{\`{a}} di Bari, via
Amendola 173, 70126, Bari, Italy\\ \vspace{0.5cm} } \vspace{1.0cm}

{\bf Abstract\\}
\end{center}

{\noindent \small  In this paper we calculate explicitly the
classical secular precessions of the node $\Omega$ and the perigee
$\omega$ of an Earth artificial satellite induced by the even
zonal harmonics of the static part of the geopotential up to
degree $l=20$. Subsequently, their systematic errors induced by
the mismodelling in the even zonal spherical harmonics
coefficients $J_l$ are compared to the general relativistic
secular gravitomagnetic and gravitoelectric precessions of the
node and the perigee of the existing laser--ranged geodetic
satellites and of the proposed LARES. The impact of the future
terrestrial gravity models from CHAMP and GRACE missions is
discussed as well. Preliminary estimates with the recently
released EIGEN--1S gravity model including the first CHAMP data
are presented.}
\end{titlepage} \newpage \pagestyle{myheadings} \setcounter{page}{1}
\vspace{0.2cm} \baselineskip 14pt

\setcounter{footnote}{0}
\setlength{\baselineskip}{1.5\baselineskip}
\renewcommand{\theequation}{\mbox{$\arabic{equation}$}}
\section{Introduction}
Recently, great efforts have been devoted to the investigation of
the possibility of measuring some tiny general relativistic
effects in the gravitational field of the Earth by analyzing the
laser--ranged data to some existing or proposed geodetic
laser--tracked (SLR) satellites.

The most famous experiment is that peformed with LAGEOS and LAGEOS
II [{\it Ciufolini et al.}, 1998] and aimed to the detection of
the gravitomagnetic Lense--Thirring drag of inertial frames [{\it
Lense and Thirring,} 1918; {\it Ciufolini and Wheeler}, 1995] in
the gravitational field of the Earth. The analysis of the orbits
of the LAGEOS satellites could allow also for an alternative
measurement of the gravitoelectric perigee advance [{\it Ciufolini
and Wheeler}, 1995] to be performed in the gravitational field of
the Earth [{\it Iorio}, 2002; {\it Iorio et al.}, 2002a].
Moreover, the possibility of including also the data from other
existing SLR satellites in these analysis is currently
investigated [{\it Iorio}, 2002]. The proposed LAGEOS--LARES
mission [{\it Ciufolini}, 1986], whose original configuration is
currently being reanalyzed [{\it Iorio et al.}, 2002b] in view of
the inclusion of more orbital elements of various SLR satellites
in the observable to be adopted, should be of great significance
for both gravitomagnetic and gravitoelectric tests [{\it Iorio et
al.}, 2002a; {\it Iorio et al.}, 2002b]. Satellite laser ranging
could be the natural candidate also for the implementation of a
space--based experiment aimed to the detection of the so called
gravitomagnetic clock effect [{\it Mashhoon et al.}, 1999; {\it
Iorio et al.}, 2002c], which is sensitive to the direction of
motion of two counter--orbiting satellites along identical orbits
in the gravitational field of a central rotating mass.

In all such performed or proposed experiments it is of the utmost
importance to reliably assess the error budget. Indeed, the
terrestrial space environment is rich of competing classical
perturbing forces of gravitational and non--gravitational origin
which in many cases are far larger than the general relativistic
effects to be investigated. In particular, it is the impact of the
systematic errors induced by the mismodelling in such various
classical perturbations which is relevant in determining the total
realistic accuracy of an experiment like those previously
mentioned.

The general relativistic effects of interest here are linear
trends affecting the perigee $\omega$ and the node $\Omega$ of the
orbit of a satellite and amounting to $10^1$--$10^3$
milliarcseconds per year (mas/y in the following) for the
gravitomagnetic and the gravitoelectric effects, respectively.

In this context the most important source of systematic error is
represented by the secular classical precessions of the node and
the perigee induced by the mismodelled even ($l=2n,\ n=1,2,3,...$)
zonal ($m=0$) harmonics $\delta J_2,\ \delta J_4,\ \delta J_6,...$
of the multipolar expansion of the terrestrial gravitational
field, called geopotential. Indeed, while the time--varying
orbital tidal perturbations [{\it Iorio}, 2001; {\it Iorio and
Pavlis}, 2001; {\it Pavlis and Iorio}, 2002] and
non--gravitational orbital perturbations [{\it Lucchesi}, 2001;
2002], according to their periods $P$ and to the adopted
observational time span $T_{\rm obs}$, can be viewed as
empirically fitted quantity and can be removed from the signal,
this is not the case of the classical even zonal secular
precessions and of certain subtle non--gravitational secular
effects of thermal origin [{\it Lucchesi}, 2002]. Their
mismodelled linear trends act as superimposed effects which may
alias the recovery of the genuine general relativistic features.
Such disturbing trends cannot be removed from the signal without
cancelling also the general relativistic signature, so that one
can only assess as more accurately as possible their impact on the
measurement. Then, the systematic error induced by the mismodelled
part of the geopotential can be viewed as a sort of unavoidable,
lower bound of the total systematic error.

In this paper we calculate explicitly, up to $l=20$, the
expressions of the coefficients of the classical secular
precessions on the node and the perigee due to the geopotential
(section 2). Their explicit, analytic form, although rather
cumbersome, may turn out to be useful in designing suitably
alternative relativistic observables which are not sensitive, at
least in part, to such classical aliasing effects [{\it Iorio and
Lichtenegger}, 2002; {\it Iorio and Lucchesi}, 2002]. In section 3
we work out the numerical values of the mismodelled precessions
for the existing SLR geodetic satellites and of the proposed LARES
and compare them to the general relativistic effects. The errors
for the spherical harmonics coefficients are those of EGM96
gravity model [{\it Lemoine et al.}, 1998] and of the recent
EIGEN--1S (see http://op.gfz-potsdam.de/champ/results/) which
includes the first data from CHAMP mission\footnote{It should be
pointed out that the values employed in the following for the
errors $\delta J_l$ by EIGEN-1S are the formal, uncalibrated
standard deviations. Moreover, EIGEN-1S is based only on the
satellites tracking data.}. These estimates can be useful in
assessing the systematic errors in various possible observables,
built up with such Keplerian orbital elements, which are sensitive
to some relativistic effects. E.g., someone could look at the
relativistic perturbations of the radial, along--track and
cross--track components of the position and velocity vectors of a
satellite; they can be obtained from suitable combinations of the
Keplerian orbital elements. The same holds also for the range and
range--rate perturbations in intersatellite tracking missions like
GRACE [{\it Cheng}, 2002]. In section 4 we review the strategy of
combining the orbital residuals of the nodes and the perigees of
different laser--ranged satellites which allows for a reduction of
the impact of the geopotential's error and yield preliminary
estimates of the errors affecting two gravitomagnetic combinations
based on the first results from EIGEN--1S. They are useful in
order to get an insight of the improvements which will take place
when the full new gravity models will become available. Section 5
is devoted to the conclusions.

In Tab. 1 we quote the orbital parameters of the existing
spherical passive geodetic laser-ranged satellites Ajisai, Stella,
Starlette, WESTPAC1, ETALON1, ETALON2, LAGEOS, LAGEOS II and of
the proposed LARES. In it $a$ is the semimajor axis, $e$ is the
eccentricity, $i$ is the inclination and $n=\sqrt{GMa^{-3}}$,
where $G$ is the Newtonian gravitational constant and $M$ is the
mass of the central body, is the Keplerian mean motion. It is
worth noting that the perigees of many of them, except for
Starlette, cannot be employed for any relativistic tests due to
the notable smallness of their eccentricities.
{\small{
\begin{table} \caption{Orbital parameters of the
existing spherical passive geodetic laser-ranged satellites and of
LARES. Aj=Ajisai, Stl=Stella, Str=Starlette, WS=WESTPAC1,
E1=ETALON1, E2=ETALON2, L1=LAGEOS, L2=LAGEOS II, LR=LARES. $a$ is
in km, $i$ in deg and $n$ in s$^{-1}$. } \label{paras}
\begin{center}
\begin{tabular}{llllllllll}
\hline   & { {Aj}} & {Stl} & {Str} & {WS} & {E1} & {E2} & L1 & L2 & LR\\
\hline
$a$  & 7,870 & 7,193 & 7,331 & 7,213 & 25,498 & 25,498 & 12,270 & 12,163 & 12,270\\
$e$ & 0.001 & 0 & 0.0204 & 0 & 0.00061 & 0.00066 & 0.0045 & 0.014 & 0.04\\
$i$  & 50 & 98.6 & 49.8 & 98 & 64.9 & 65.5 & 110 & 52.65 & 70\\
$n$  & 0.0009 & 0.001 & 0.001 & 0.001 & 0.00015 & 0.00015 &
0.00046 & 0.00047 &
0.00046\\
\hline
\end{tabular}
\end{center}
\end{table}
}}

\section{The orbital classical precessions}
Here we show the explicitly calculated coefficients \eqi\djn\eqf
and \eqi\djo\eqf of the satellites' classical nodal and apsidal
precessions due to the even zonal harmonics of the geopotential up
to $l= 20$. The classical precessions of the node and the perigee
due to the even zonal harmonics of geopotential can be written as
\begin{eqnarray}
\dot\Omega^{\rm even\ zonal}_{\rm class}&=&\sum_{n=1}\dot\Omega_{.2n}\times J_{2n},\\
\dot\omega^{\rm even\ zonal}_{\rm
class}&=&\sum_{n=1}\dot\omega_{.2n}\times J_{2n}.
\end{eqnarray} As we shall see later, the coefficients
$\dot\Omega_{.2n}$ and $\dot\omega_{.2n}$ depend only on the
orbital parameters of the satellites. Recall that $J_{l}\ \equiv
-C_{l0},\ l=2n,\ n=1,2,3...$ where the unnormalized adimensional
{Stokes} coefficients $C_{lm}$ of degree $l$ and order $m$ can be
obtained from the normalized ${\overline{C}}_{lm}$ with \eqi
C_{lm}=N_{lm}{\overline{C}}_{lm}. \eqf In it \eqi N_{lm}=
\left[\frac{(2l+1)(2-\delta_{0m})(l-m)!}{(l+m)!}\right]^{\frac{1}{2}}.\eqf
The general expressions of the classical rates of the near {Earth}
satellites' Keplerian orbital elements due to the {geopotential}
${\dot a}_{\rm class},\ {\dot e}_{\rm class},\ {\dot i}_{\rm
class},\ {\dot \Omega}_{\rm class},\ {\dot \omega}_{\rm class},\
{\dot {{\mathcal{M}}}}_{\rm class}$, and of the inclination
functions $F_{lmp}(i)$ and of the eccentricity functions
$G_{lpq}(e)$ can be found in [{\it {Kaula}}, 1966]. The
coefficients $\dot\Omega_{.2n}$ and $\dot\omega_{.2n}$ are of
crucial importance in the evaluation of the systematic error due
to the mismodelled even zonal harmonics of the {geopotential};
moreover, they enter the combined residuals' coefficients $c_i$
about which we speak in section 4. Since the \grl\ effects
investigated are secular perturbations, we have considered only
the perturbations averaged over one satellite' s orbital period.
This has been accomplished with the condition $l-2p+q = 0$ which
allows for canceling out the rate of the mean anomaly
${\mathcal{M}}$. Since the {eccentricity} functions $G_{lpq}$ are
proportional to $e^{\vass{q}}$, for a given value of $l$ we have
considered only those values of $p$ which fulfil the condition
$l-2p+q = 0$ with $q=0$, i.e. $p=\frac{l}{2}$. This implies that
in the summations
\eqi\sum_{p=0}^{l}\dert{F_{l0p}}{i}\sum_{q=-\infty}^{+\infty}G_{lpq}\eqf
and
\eqi\sum_{p=0}^{l}F_{l0p}\sum_{q=-\infty}^{+\infty}\dert{G_{lpq}}{e}\eqf
involved in the expressions of the classical rates we have
considered only $F_{l0\frac{l}{2}}$ and $G_{l\frac{l}{2}0}$.
Moreover, in working out the $G_{l\frac{l}{2}0}$ we have neglected
the terms of order ${\mathcal{O}}(e^{k})$ with $k>2$.

\subsection{The nodal coefficients}

The nodal coefficients, proportional to \eqi\frac{1}{\sin
i}\sum_{q=-\infty}^{+\infty}G_{lpq}\sum_{p=0}^{l}\dert{F_{lmp}}{i},\eqf
are ($R$ is the Earth's mean equatorial radius)

\eqia\jn2 & = &  -\frac{3}{2}n\rtr{2}\frac{\cos i}{\ecn{2}},\\
\jn{4} & = & \jn2\left[ \frac{5}{8}\rtr{2}
\frac{(1+\frac{3}{2}e^{2})}{\ecn{2}} \left(
7\sus{2}-4\right)\right],\\
\jn{6} & = & \jn2\left[ \frac{35}{8}\rtr{4}
\frac{(1+5e^{2})}{\ecn{4}} \left(
\frac{33}{8}\sus{4}-\frac{9}{2}\sus{2}+1\right)\right],\\
\jn{8} & = & \jn2\left[ \frac{105}{16}\rtr{6}
\frac{(1+\frac{21}{2}e^{2})}{\ecn{6}} \left( \frac{715}{64}\sus{6}
-\frac{143}{8}\sus{4}+\right.\right.\nonumber\\
&+&\left.\left.\frac{33}{4}\sus{2}-1\right)\right] ,\\
\jn{10} & = & \jn2\left[  \frac{1,155}{128} \rtr{8}
\frac{(1+18e^{2})}{\ecn{8}}\left(
\frac{4,199}{128}\sus{8}-\frac{1,105}{16}\sus{6}\right.\right.\nonumber\\\nonumber\\
& +&
\left.\left.\frac{195}{4}\sus{4}-13\sus{2}+1\right)\right],\\
\jn{12} & = & \jn2\left[ \frac{3,003}{256}\rtr{10}
\frac{\ekn{55}{2}}{\ecn{10}} \left( \frac{52,003}{512}\sus{10}-
\frac{33,915}{128}\sus{8}\right.\right.\nonumber\\\nonumber\\
&+ &  \left.\left.
\frac{8,075}{32}\sus{6}-\frac{425}{4}\sus{4}+\frac{75}{4}\sus{2}-1\right)\right],\\
\jn{14} & = & \jn2\left[ \frac{15,015}{1,024}\rtr{12}
\frac{\ekn{91}{2}}{\ecn{12}} \left(
\frac{334,305}{1,024}\sus{12}-\frac{260,015}{256}\sus{10}\right.\right.\nonumber\\\nonumber\\
& +& \left.\left.
\frac{156,009}{128}\sus{8}-\frac{11,305}{16}\sus{6}+\frac{1,615}{8}\sus{4}-\frac{51}{2}\sus{2}+1\right)\right],\\
\jn{16} & = & \jn2\left[ \frac{36,465}{2,048}\rtr{14}
\frac{\ekn{105}{2}}{\ecn{14}} \left(
\frac{17,678,835}{16,384}\sus{14}-\right.\right.\nonumber\\\nonumber\\
&-&\left.\left.\frac{3,991,995}{1,024}\sus{12} +
\frac{2,890,755}{512}\sus{10}-
\frac{535,325}{128}\sus{8}+\frac{107,065}{64}\sus{6}\right.\right.\nonumber\\\nonumber\\
&-&
\left.\left.\frac{2,793}{8}\sus{4}+\frac{133}{4}\sus{2}-1\right)\right],\\
\jn{18} & = & \jn2\left[ \frac{692,835}{32,768}\rtr{16}
\frac{(1+68e^{2})}{\ecn{16}}
\left(\frac{119,409,675}{32,768}\sus{16}-\right.\right.\nonumber\\\nonumber\\&-&
\frac{30,705,345}{2,048}\sus{14}
+\frac{6,513,255}{256}\sus{12}-\frac{1,470,735}{64}\sus{10}+\nonumber\\\nonumber\\
&+&\left.\left.\frac{760,725}{64}\sus{8} -
\frac{28,175}{8}\sus{6}+\right.\right.\nonumber\\\nonumber\\
&+&\left.\left.\frac{1,127}{2}\sus{4}-
42\sus{2}+1\right)\right],\\
\jn{20} & = & \jn2\left[ \frac{1,616,615}{65,536}\rtr{18}
\frac{\ekn{171}{2}}{\ecn{18}}
\left(\frac{1,641,030,105}{131,072}\sus{18}\right.\right.\nonumber\\\nonumber\\
& -&\frac{1,893,496,275}{32,768}\sus{16}+
\frac{460,580,175}{4,096}\sus{14}-\frac{30,705,345}{256}\sus{12}\nonumber\\\nonumber\\
& +& \frac{19,539,765}{256}\sus{10}
-\frac{1,890,945}{64}\sus{8}+\frac{108,675}{16}\sus{6}-\nonumber\\\nonumber\\
&-&\left.\left.\frac{1,725}{2}\sus{4} +
\frac{207}{4}\sus{2}-1\right)\right]. \eqfa

\subsection{The perigee coefficients}

The coefficients of the classical perigee precession are much more
involved because they are proportional to \eqi-\left(\frac{\cos
i}{\sin
i}\right)\sum_{q=-\infty}^{+\infty}G_{lpq}\sum_{p=0}^{l}\dert{F_{lmp}}{i}+
\frac{(1-e^2)}{e}\sum_{q=-\infty}^{+\infty}\dert{G_{lpq}}{e}\sum_{p=0}^{l}F_{lmp}.\eqf
We can pose  $\jo{2n}=\jo{2n}^{a}+\jo{2n}^{b}$.

The first set is given by ($R$ is the Earth's mean equatorial
radius)

\eqia\jo2^{a} & = &  \frac{3}{2}n\rtr{2}\frac{\cos^{2}
i}{\ecn{2}},\\
\jo{4}^{a} & = & \jo2^{a}\left[ \frac{5}{8}\rtr{2}
\frac{(1+\frac{3}{2}e^{2})}{\ecn{2}} \left(
7\sus{2}-4\right)\right],\\
\jo{6}^{a} & = & \jo2^{a}\left[ \frac{35}{8}\rtr{4}
\frac{(1+5e^{2})}{\ecn{4}} \left(
\frac{33}{8}\sus{4}-\frac{9}{2}\sus{2}+1\right)\right],\\
\jo{8}^{a} & = & \jo2^{a}\left[ \frac{105}{16}\rtr{6}
\frac{(1+\frac{21}{2}e^{2})}{\ecn{6}} \left( \frac{715}{64}\sus{6}-\right.\right.\nonumber\\\nonumber\\
&-&\left.\left.\frac{143}{8}\sus{4}+\frac{33}{4}\sus{2}-1\right)\right] ,\\
\jo{10}^{a} & = & \jo2^{a}\left[  \frac{1,155}{128} \rtr{8}
\frac{(1+18e^{2})}{\ecn{8}}\left(
\frac{4,199}{128}\sus{8}-\frac{1,105}{16}\sus{6}\right.\right.\nonumber\\\nonumber\\
& +&
\left.\left.\frac{195}{4}\sus{4}-13\sus{2}+1\right)\right],\\
\jo{12}^{a} & = & \jo2^{a}\left[ \frac{3,003}{256}\rtr{10}
\frac{\ekn{55}{2}}{\ecn{10}} \left( \frac{52,003}{512}\sus{10}-
\frac{33,915}{128}\sus{8}\right.\right.\nonumber\\\nonumber\\
&+ &  \left.\left.
\frac{8,075}{32}\sus{6}-\frac{425}{4}\sus{4}+\frac{75}{4}\sus{2}-1\right)\right],\\
\jo{14}^{a} & = & \jo2^{a}\left[ \frac{15,015}{1,024}\rtr{12}
\frac{\ekn{91}{2}}{\ecn{12}} \left(
\frac{334,305}{1,024}\sus{12}-\frac{260,015}{256}\sus{10}+\right.\right.\nonumber\\\nonumber\\
& +& \left.\left.
\frac{156,009}{128}\sus{8}-\frac{11,305}{16}\sus{6}+\frac{1,615}{8}\sus{4}-\right.\right.\nonumber\\\nonumber\\
&-&\left.\left.\frac{51}{2}\sus{2}+1\right)\right],\\
\jo{16}^{a} & = & \jo2^{a}\left[ \frac{36,465}{2,048}\rtr{14}
\frac{\ekn{105}{2}}{\ecn{14}} \left(
\frac{17,678,835}{16,384}\sus{14}-\frac{3,991,995}{1,024}\sus{12}\right.\right.\nonumber\\\nonumber\\
&+ &  \left.\left.\frac{2,890,755}{512}\sus{10}-
\frac{535,325}{128}\sus{8}+\frac{107,065}{64}\sus{6}\right.\right.\nonumber\\\nonumber\\
&- &
\left.\left.\frac{2,793}{8}\sus{4}+\frac{133}{4}\sus{2}-1\right)\right],\\
\jo{18}^{a} & = & \jo2^{a}\left[ \frac{692,835}{32,768}\rtr{16}
\frac{(1+68e^{2})}{\ecn{16}}
\left(\frac{119,409,675}{32,768}\sus{16}-
\right.\right.\nonumber\\\nonumber\\
&-&\frac{30,705,345}{2,048}\sus{14}+\frac{6,513,255}{256}\sus{12}-\frac{1,470,735}{64}\sus{10}+
\nonumber\\\nonumber\\
&+&\left.\left.\frac{760,725}{64}\sus{8}-\frac{28,175}{8}\sus{6}+\frac{1,127}{2}\sus{4}-\right.\right.\nonumber\\\nonumber\\
&-&\left.\left.42\sus{2}+1\right)\right],\\
\jo{20}^{a} & = & \jo2^{a}\left[ \frac{1,616,615}{65,536}\rtr{18}
\frac{\ekn{171}{2}}{\ecn{18}}
\left(\frac{1,641,030,105}{131,072}\sus{18}\right.\right.\nonumber\\\nonumber\\
& -&\frac{1,893,496,275}{32,768}\sus{16}+
\frac{460,580,175}{4,096}\sus{14}-\frac{30,705,345}{256}\sus{12}\nonumber\\\nonumber\\
& +& \left.\left.\frac{19,539,765}{256}\sus{10}
-\frac{1,890,945}{64}\sus{8}+\frac{108,675}{16}\sus{6}-\right.\right.\nonumber\\\nonumber\\
&-&\left.\left.\frac{1,725}{2}\sus{4}+\frac{207}{4}\sus{2}-1\right)\right].
\eqfa The second set is given by ($R$ is the Earth's mean
equatorial radius)
\eqia \jw2 & = & -\frac{3}{2}n\rtr2,\\
\jo2^{b} & = &
\jw2\left\{\left[\frac{1}{\ecn{2}}\right]\left(\frac{3
}{2}\sin^2 i-1\right)\right\},\\
\jo4^{b} & = & \jw2\left\{\frac{5}{8}\rtr2\left[
\frac{3}{\ecn{3}}+7\frac{\ekn{3}{2}}{\ecn{4}}\right]\left(\frac{7}{4}\sus4-\right.\right.\nonumber\\
&-&\left.\left.2\sus2+\frac{2}{5}\right)\right\},\\
\jo{6}^{b} & = & \jw2 \left\{\frac{35}{8}\rtr{4}\left[
\frac{10}{\ecn{5}}+11\frac{(1+5e^2)}{\ecn{6}}\right]\left(\frac{33}{48}\sus{6}
\right.\right.\nonumber\\\nonumber\\
&-& \left.\left.\frac{9}{8}\sus{4}+\frac{1}{2}\sus{2}-\frac{1}{21}
\right)\right\},\\
\jo{8}^{b} & = & \jw2\left\{\frac{105}{16}\rtr{6}\left[
\frac{21}{\ecn{7}}+15\frac{\ekn{21}{2}}{\ecn{8}}\right]\left(\frac{715}{512}\sus{8}
\right.\right.\nonumber\\\nonumber\\
&-&
\left.\left.\frac{143}{48}\sus{6}+\frac{33}{16}\sus{4}-\frac{1}{2}\sus{2}
+\frac{1}{36}
\right)\right\},\\
\jo{10}^{b} & = & \jw2 \left\{\frac{1,155}{128}\rtr{8}\left[
\frac{36}{\ecn{9}}+19\frac{(1+18e^2)}{\ecn{10}}\right]\left(\frac{4,199}{1,280}\sus{10}
\right.\right.\nonumber\\\nonumber\\
&-&
\left.\left.\frac{1,105}{128}\sus{8}+\frac{195}{24}\sus{6}-\frac{13}{4}\sus{4}
+\frac{1}{2}\sus{2}-\frac{1}{55}
\right)\right\},\\
\jo{12}^{b} & = & \jw2 \left\{\frac{3,003}{256}\rtr{10}\left[
\frac{55}{\ecn{11}}+23\frac{\ekn{55}{2}}{\ecn{12}}\right]\left(\frac{52,003}{6,144}\sus{12}
\right.\right.\nonumber\\\nonumber\\
&-&
\frac{6,783}{256}\sus{10}+\frac{8,075}{256}\sus{8}-\frac{425}{24}\sus{6}
+\frac{75}{16}\sus{4}\nonumber\\\nonumber\\
&-&\left.\left.\frac{1}{2}\sus{2}+\frac{1}{78}
\right)\right\},\\
\jo{14}^{b} & = & \jw2 \left\{\frac{15,015}{1,024}\rtr{12}\left[
\frac{91}{\ecn{13}}+27\frac{\ekn{91}{2}}{\ecn{14}}\right]\right.\times\nonumber\\\nonumber\\
&\times&\left.\left(\frac{334,305}{14,336}\sus{14}-
\frac{260,015}{3,072}\sus{12}+\frac{156,009}{1,280}\sus{10}-\right.\right.\nonumber\\\nonumber\\
&-&\left.\left.\frac{11,305}{128}\sus{8}+\frac{1,615}{48}\sus{6}-\frac{51}{8}\sus{4}+\right.\right.\nonumber\\\nonumber\\
&+&\left.\left.\frac{1}{2}\sus{2}-\frac{1}{105}
\right)\right\},\\
\jo{16}^{b} & = & \jw2\left\{ \frac{36,465}{2,048}\rtr{14}\left[
\frac{105}{\ecn{15}}+31\frac{\ekn{105}{2}}{\ecn{16}}\right]\right.\times\nonumber\\\nonumber\\
&\times&\left.\left(\frac{17,678,835}{262,144}\sus{16}-
\frac{570,285}{2,048}\sus{14}+\frac{963,585}{2,048}\sus{12}-\right.\right.\nonumber\\\nonumber\\
&-&\left.\left.\frac{107,065}{256}\sus{10}
+\frac{107,065}{512}\sus{8}
-\frac{931}{16}\sus{6}+\right.\right.\nonumber\\\nonumber\\
&+&\left.\left.\frac{133}{16}\sus{4}-\frac{1}{2}\sus{2}
+\frac{1}{136}
\right)\right\},\\
\jo{18}^{b} & = & \jw2\left\{ \frac{692,835}{32,768}\rtr{16}\left[
\frac{136}{\ecn{17}}+35\frac{(1+68e^2)}{\ecn{18}}\right]\right.\times\nonumber\\\nonumber\\
&\times&\left.\left(\frac{39,803,225}{196,608}\sus{18}-
\frac{30,705,345}{32,768}\sus{16}+\frac{930,465}{512}\sus{14}-\right.\right.\nonumber\\\nonumber\\
&-&\left.\left.\frac{490,245}{256}\sus{12} +
\frac{152,145}{128}\sus{10} -
\frac{28,175}{64}\sus{8}+\right.\right.\nonumber\\\nonumber\\
&+&\left.\left.\frac{1,127}{12}\sus{6}-\frac{21}{2}\sus{4}
+\frac{1}{2}\sus{2}-\frac{1}{171}
\right)\right\},\\
\jo{20}^{b} & = & \jw2\left\{
\frac{1,616,615}{65,536}\rtr{18}\left[
\frac{171}{\ecn{19}}+39\frac{\ekn{171}{2}}{\ecn{20}}\right]\right.\times\nonumber\\\nonumber\\
&\times&\left.\left(\frac{328,206,021}{524,288}\sus{20}-
\frac{210,388,475}{65,536}\sus{18}+\frac{460,580,175}{65,536}\sus{16}-\right.\right.\nonumber\\\nonumber\\
&-&\left.\left.\frac{30,705,345}{3,584}\sus{14}+
\frac{6,513,255}{1,024}\sus{12}-\frac{378,189}{128}\sus{10}+\right.\right.\nonumber\\\nonumber\\
&+&\left.\left.\frac{108,675}{128}\sus{8}-\frac{575}{4}\sus{6}+\frac{207}{16}\sus{4}-\right.\right.\nonumber\\\nonumber\\
&-&\left.\left.\frac{1}{2}\sus2+ \frac{1}{210} \right)\right\}.
\eqfa

\section{The mismodelled classical precessions}
The results obtained in the previous section can be used in
working out explicitly the contributions of the mismodelled
classical nodal and apsidal precessions up to degree $l=20$ of the
existing spherical passive laser-ranged geodetic satellites and of
the proposed LARES. They are of the form $\delta\dot
\Omega_{(2n)}=\dot \Omega _{.2n}\times \delta J_{2n}$,
$n=1,2,...10$ and $\delta\dot \omega_{(2n)}=\dot \omega
_{.2n}\times \delta J_{2n}$, $n=1,2,...10$. The coefficients $\dot
\Omega _{.2n}$ and $\dot \omega _{.2n}$ are worked out in section
2 and the values employed for $\delta
J_{2n}=-\sqrt{4n+1}\times\delta{\overline{C}}_{2n\ 0}$,
$n=1,2,...10$ are those quoted in the adopted Earth's gravity
model.
\begin{table}[htb]
\caption{Mismodelled classical nodal precessions
$\delta\dot\Omega_{(2n)}$ and predicted Lense-Thirring nodal
precessions $\dot\Omega_{\rm LT}$ of the existing geodetic
laser-ranged satellites and of LARES. L1={LAGEOS}, L2={LAGEOS II},
LR={LARES}, Aj={Ajisai}, Stl={Stella}, Str={Starlette},
WS={WESTPAC1}, E1={ETALON1}, E2={ETALON2}. All the values are in
mas/y. For the ETALON satellites, when the values are less than
$10^{-4}$ mas/y a -- has been inserted. EGM96 gravity model has
been adopted. }\label{nodotab}
\begin{center}
\begin{tabular}{llllllllll}
\hline $2n$ & {L1} & { L2} & {LR} &
{Aj} & {Stl }& {Str}& { WS}& { E1} & {E2}\\
\hline
2 & -33.4 & 61 & 33.4 & 296.8 & -94.6 & 382.3 & -87.1 & 3.2 & 3.1\\
4 & -48.3 & 17.4 & 48.7 & 51.5 & -519 & 59.5 & -479.2 & 0.8 & 0.8\\
6 & -17 & -26.1 & 17.3 & -809.7 & -912.2 & -1,397.7 & -847.9 & 0.03 & 0.03\\
8 & -1.9 & -10.3 & 2 & -366.3 & -1,487.2 & -674.4 & -1,399.7 & -0.005 & -0.004\\
10 & 2.1 & 3.1 & -2.2 & 823.5 & -1,855 & 1,933.4 & -1,781.8 & -0.001 & --\\
12 & 1.6 & 2.5 & -1.7 & 647.5 & -2,144.6 & 1,636.4 & -2,126.6 & -- & --\\
14 & 0.6 & -0.007 & -0.6 & -542.6 & -1,963.4 & -1,780.9 & -2,049.4 & -- & --\\
16 & 0.09 & -0.2 & -0.1 & -517.2 & -1,204.6 & -1,787.9 & -1,376.8 & -- & --\\
18 & -0.007 & -0.03 & 0.008 & 117.9 & -512.4 & 580 & -717 & -- & --\\
20 & -0.01 & 0.01 & 0.01 & 247.6 & -79.5 & 1,177 & -309 & -- & --\\
\hline
$\dot\Omega_{\rm LT}$ & 30.7 & 31.6 & 30.8 & 116.7 & 152.8 & 144.4 & 151.5 & 3.4 & 3.4\\
\hline
\end{tabular}
\end{center}
\end{table}



From Tab. 2 it is interesting to note that for the satellites
orbiting at lower altitudes than the LAGEOS satellites the impact
of the mismodelled part of the geopotential does not reduce to the
first two or three even zonal harmonics. This feature is very
important in calculating the error budget, especially if the nodes
of low--orbiting satellites are to be considered. Moreover, while
for the LAGEOS family a calculation up to $l=20$ is rather
adequate, this is not the case for the other satellites for which
the even zonal harmonics of degree $l>20$ should be considered as
well. In regard to this topic, the choice of the Earth gravity
model becomes crucial because EGM96, for example, does not seem
to be particularly reliable at degrees higher than 20. The same
considerations hold also for the perigee whose mismodelled
classical precessions are quoted in Tab. 3. We have considered
only LAGEOS II, Starlette and the LARES due to the extreme
smallness of the eccentricity of the other satellites.
\begin{table}[htb]
\caption{Mismodelled classical perigee precessions
$\delta\dot\omega_{(2n)}$ and predicted Lense-Thirring and
gravitoelectric perigee precessions $\dot\omega_{\rm LT}$ and
$\dot\omega_{\rm GE}$ of the existing spherical passive geodetic
laser-ranged satellites and of LARES. L1=LAGEOS, L2=LAGEOS II,
LR=LARES, Aj=Ajisai, Stl=Stella, Str=Starlette, WS=WESTPAC1,
E1=ETALON1, E2=ETALON2. All the values are in mas/y. EGM96 gravity
model has been adopted}\label{peritab}
\begin{center}
\begin{tabular}{llllllllll}
\hline $2n$ & L1 & L2 & LR&
Aj & Stl & Str & WS & E1 & E2\\
\hline
2 & -- & -42.3 & 20.3 & -- & -- & -320.7 & -- & -- & --\\
4 & -- & -122.7 & -17.6 & -- & -- & -1,924.4 & -- & -- & --\\
6 & -- & -18.2 & -49.2 & -- & -- & 429.1 & -- & -- & --\\
8 & -- & 43.1 & -42.6 & -- & -- & 6,355.8 & -- & -- & --\\
10 & -- & 19.5 & -18 & -- & -- & 2,805.1 & -- & -- & --\\
12 & -- & -5.3 & -3 & -- & -- & -10,862.2 & -- & -- & --\\
14 & -- & -6.2 & 2 & -- & -- & -10,774.7 & -- & -- & --\\
16 & -- & -0.2 & 1.3 & -- & -- & 8,395.8 & -- & --& --\\
18 & -- & 0.4 & 0.4 & -- & -- & 9,086.4 & -- &-- & --\\
20 & -- & 0.1 & 0.08 & -- & -- & -3,043.3 & -- & -- & --\\
\hline
$\dot\omega_{\rm LT}$ & --& -57.5 & -31.6 & -- & -- & 68.5 & -- & -- & --\\
\hline
$\dot\omega_{\rm GE}$ & -- & 3,348 & 3,278.6 & -- & -- & 11,804.7 & -- & -- & -- \\
\hline
\end{tabular}
\end{center}
\end{table}
\begin{table}[htb]
\caption{Mismodelled classical nodal precessions
$\delta\dot\Omega_{(2n)}$ and predicted Lense-Thirring nodal
precessions $\dot\Omega_{\rm LT}$ of the existing geodetic
laser-ranged satellites and of LARES. L1={LAGEOS}, L2={LAGEOS II},
LR={LARES}, Aj={Ajisai}, Stl={Stella}, Str={Starlette},
WS={WESTPAC1}, E1={ETALON1}, E2={ETALON2}. The errors $\delta
J_{2n}$ are those of the preliminary EIGEN--1S Earth gravity model
from 88 days of CHAMP data. All the values are in mas/y. For the
ETALON satellites, since all the values are less than $10^{-1}$
mas/y a -- has been inserted.}\label{nodotab2}
\begin{center}
\begin{tabular}{llllllllll}
\hline $2n$ & {L1} & { L2} & {LR} &
{Aj} & {Stl }& {Str}& { WS}& { E1} & {E2}\\
\hline
2 & -3.9 & 7.2 & 3.9 & 35 & -11.1 & 45.1 & -10.2 & -- & --\\
4 & -7.2 & 2.6 & 7.3 & 7.7 & -77.7 & 8.9 & -71.7 & -- & --\\
6 & -3.8 & -5.8 & 3.8 & -181.5 & -204.5 & -313.4 & -190.1 & -- & --\\
8 & -0.4 & -2.3 & 0.4 & -83.8 & -340.3 & -154.3 & -320.3 & -- & --\\
10 & 0.4 & 0.6 & -0.4 & 168.5 & -379.6 & 395.6 & -364.6 & --- & --\\
12 & 0.3 & 0.4 & -0.3 & 125.9 & -417 & 318.2 & -413.5 & -- & --\\
14 & 0.1 & -0.001 & -0.1 & -122.1 & -441.9 & -400.8 & -461.3 & -- & --\\
16 & 0.02 & -0.07 & -0.03 & -152 & -354.1 & -525.6 & -404.7 & -- & --\\
18 & -0.003 & -0.01 & 0.003 & 49.2 & -214 & 242.2 & -299.4 & -- & --\\
20 & -0.005 & 0.007 & 0.006 & 131.3 & -42.2 & 624.3 & -163.9 & -- & --\\
\hline
$\dot\Omega_{\rm LT}$ & 30.7 & 31.6 & 30.8 & 116.7 & 152.8 & 144.4 & 151.5 & 3.4 & 3.4\\
\hline
\end{tabular}
\end{center}
\end{table}
\begin{table}[htb]
\caption{Mismodelled classical perigee precessions
$\delta\dot\omega_{(2n)}$ and predicted Lense-Thirring and
gravitoelectric perigee precessions $\dot\omega_{\rm LT}$ and
$\dot\omega_{\rm GE}$ of the existing spherical passive geodetic
laser-ranged satellites and of LARES. L1=LAGEOS, L2=LAGEOS II,
LR=LARES, Aj=Ajisai, Stl=Stella, Str=Starlette, WS=WESTPAC1,
E1=ETALON1, E2=ETALON2. All the values are in mas/y. The errors
$\delta J_{2n}$ are those of the preliminary EIGEN--1S Earth
gravity model from 88 days of CHAMP data. }\label{peritab2}
\begin{center}
\begin{tabular}{llllllllll}
\hline $2n$ & L1 & L2 & LR&
Aj & Stl & Str & WS & E1 & E2\\
\hline
2 & -- & -4.9 & 2.4 & -- & -- & -37.8 & -- & -- & --\\
4 & -- & -18.3 & -2.6 & -- & -- & -288.2 & -- & -- & --\\
6 & -- & -4 & -11 & -- & -- & 96.2 & -- & -- & --\\
8 & -- & 9.8 & -9.7 & -- & -- & 1,454.6 & -- & -- & --\\
10 & -- & 4 & -3.6 & -- & -- & 574 & -- & -- & --\\
12 & -- & -1 & -0.6 & -- & -- & -2,112.2 & -- & -- & --\\
14 & -- & -1.4 & 0.4 & -- & -- & -2,425.3 & -- & -- & --\\
16 & -- & -0.07 & 0.3 & -- & -- & 2,468.3 & -- & --& --\\
18 & -- & 0.2 & 0.1 & -- & -- & 3,794.6 & -- &-- & --\\
20 & -- & 0.05 & 0.04 & -- & -- & -1,614.6 & -- & -- & --\\
\hline
$\dot\omega_{\rm LT}$ & --& -57.5 & -31.6 & -- & -- & 68.5 & -- & -- & --\\
\hline
$\dot\omega_{\rm GE}$ & -- & 3,348 & 3,278.6 & -- & -- & 11,804.7 & -- & -- & -- \\
\hline
\end{tabular}
\end{center}
\end{table}
From both Tab. 2 and Tab. 3 the relevant impact of the first two
or three even zonal harmonics, at least for the LAGEOS satellites,
is quite apparent. The situation for the lower orbiting satellites
is far more unfavorable.

In Tab. 4 and Tab. 5 we repeat the analysis with the very recently
released EIGEN--1S gravity model which includes the first data
from CHAMP.

From an inspection of Tab. 2--Tab. 5 it turns out very clearly
that, if the orbital elements of satellites other than those of
LAGEOS family are to be considered, the observable which would
account for them must cope with the problem of reducing the impact
also of the degrees higher than 4.

\section{The systematic zonal error}
A possible strategy for reducing the impact of the geopotential's
error consists of suitable combinations of the orbital residuals
of the rates of the nodes and the perigees of different SLR
satellites [{\it Ciufolini}, 1996; {\it Iorio,} 2002]. Such
combinations can be written in the form \eqi\sum_{i=1}^N c_i
f_i=X_{\rm GR}\mu_{\rm GR},\lb{combi}\eqf in which the
coefficients $c_i$ are, in general, suitably built up with the
orbital parameters of the satellites entering the combinations,
the $f_i$ are the residuals of the rates of the nodes and the
perigees of the satellites entering the combination, $X_{\rm GR}$
is the slope, in mas/y, of the general relativistic trend of
interest and $\mu_{\rm GR}$ is the solve--for parameter, to be
determined by means of usual least--square procedures, which
accounts for the general relativistic effect. For example, in the
case of the Lense--Thirring--LAGEOS experiment [{\it Ciufolini},
1996] $X_{\rm LT}=60.2$ mas/y, while for the gravitoelectric
perigee advance [{\it Iorio}, 2002] $X_{\rm GE}=3,348$ mas/y. More
precisely, the combinations of \rfr{combi} are obtained in the
following way. The equations for the residuals of the rates of the
$N$ chosen orbital elements are written down, so to obtain a non
homogeneous algebraic linear system of $N$ equations in $N$
unknowns. They are $\mu_{\rm GR}$ and the first $N-1$ mismodelled
spherical harmonics coefficients $\delta J_l$ in terms of which
the residual rates are expressed. The coefficients $c_i$ and,
consequently, $X_{\rm GR}$ are obtained by solving for $\mu_{\rm
GR}$ the system of equations. So, the coefficients $c_i$ are
calculated in order to cancel out the contributions of the first
$N-1$ even zonal mismodelled harmonics which, as we have seen in
the previous section, represent the major source of uncertainty in
the Lense--Thirring and gravitoelectric precessions [{\it
Ciufolini}, 1996; {\it Iorio}, 2002]. The coefficients $c_i$ can
be either constant \footnote{In general, the coefficient of the
first orbital element entering a given combination is equal to 1,
as for the combinations in [{\it Ciufolini}, 1996; {\it Iorio},
2002]. } or depend on the orbital elements of the satellites
entering the combinations through the coefficients
$\dot\Omega_{.2n}$ and $\dot\omega_{.2n}$ worked out in section 2.

Now we expose how to calculate the systematic error due to the
mismodelled even zonal harmonics of the geopotential for the
combinations involving the residuals of the nodes and the perigees
of various satellites.

In general, if we have an observable $q$ which is a function
$q=q(x_j)$, $j=1,2...M$ of $M$ correlated parameters $x_j$ the
error in it is given by
 \eqi \delta
q=\left[\sum_{j=1}^M\left(\derp{q}{x_j}\right)^{2}\sigma_j^2+2\sum_{h\neq
k
=1}^M\left(\derp{q}{x_h}\right)\left(\derp{q}{x_k}\right)\sigma^{2}_{hk}\right]^{\frac{1}{2}}\lb{app1}\eqf
in which $\sigma^{2}_{j}\equiv C_{jj}$ and $\sigma^{2}_{hk}\equiv
C_{hk}$ where $\{C_{hk}\}$ is the square matrix of covariance of
the parameters $x_j$.

In our case the observable $q$ is any residuals' combination \eqi
q=\sum_{i=1}^{N}c_i f_i(x_j),\ j=1,2...10,\eqf where $x_j,\
j=1,2...10$ are the even zonal geopotential's coefficients $J_2,\
J_4...J_{20}$.  Since \eqi \derp{q}{x_j}=\sum_{i=1}^N c_i
\derp{f_i}{x_j},\ j=1,2...10\lb{app2},\eqf by putting \rfr{app2}
in \rfr{app1} one obtains, in mas/y \eqi \delta
q=\left[\sum_{j=1}^{10}\left(\sum_{i=1}^N c_i
\derp{f_i}{x_j}\right)^{2}\sigma_j^2+2\sum_{h\neq k
=1}^{10}\left(\sum_{i=1}^N c_i
\derp{f_i}{x_h}\right)\left(\sum_{i=1}^N c_i
\derp{f_i}{x_k}\right)\sigma^{2}_{hk}\right]^{\frac{1}{2}}.\lb{app3}\eqf
The percent error, for a given \grl\ trend and for a given
combination, is obtained by taking the ratio of \rfr{app3} to the
slope in mas/y of the \grl\ trend for the residual combination
considered.

The validity of \rfr{app3} has been checked by calculating with it
and the covariance matrix of EGM96 gravity model the systematic
error due to the even zonal harmonics of the geopotential of the
gravitomagnetic LAGEOS experiment; indeed the result
\eqi\delta\mu_{\rm LT}=12.9\%\ \mu_{\rm LT}\eqf claimed in [{\it
Ciufolini et al.}, 1998] has been obtained again. For the
EGM96--induced systematic error due to the even zonal harmonics of
the geopotential of alternative proposed gravitomagnetic and
gravitoelectric experiments, see [{\it Iorio}, 2002; {\it Iorio et
al.}, 2002a].

A very important point to stress is that the forthcoming new data
on the Earth's gravitational field by CHAMP, which has been
launched in July 2000, and GRACE, which has been launched in March
2002, will have a great impact on the reduction of the systematic
error due to the mismodelled part of geopotential.

In order to get a preliminary insight of what the improvement due
to the new Earth gravity models might be, let us consider, for
example, the usual observable by Ciufolini [{\it Ciufolini}, 1996]
for the detection of the Lense--Thirring drag
\eqi\delta\dot\Omega^{\rm L1}+0.295\times\delta\dot\Omega^{\rm
L2}-0.35\times\delta\dot\omega^{\rm L2}\sim 60.2\mu_{\rm LT}.\eqf
The root--sum--square error due to geopotential, according to the
diagonal part only of the covariance matrix of EGM96 model,
amounts to 46.5$\%$ \footnote{It reduces to 12.9$\%$ by
considering also the correlation among the spherical harmonics
coefficients, according to EGM96, as stated before.}; according to
the diagonal part only of the covariance matrix of EIGEN--1S (at
present, its full covariance matrix is not yet publicly
available), it reduces to 10.5$\%$.

If we consider the combination proposed in [{\it Iorio}, 2002]
\eqi\delta\dot\Omega^{\rm L1}+0.444\times\delta\dot\Omega^{\rm
L2}-0.027\times\delta\dot\Omega^{\rm
Aj}-0.341\times\delta\dot\omega^{\rm L2}\sim 61.2\mu_{\rm LT},\eqf
in this case the root--sum--square error due to geopotential,
according to the diagonal part only of the covariance matrix of
EGM96 model, amounts to 64.2$\%$ \footnote{It reduces to 10.8$\%$
by considering also the correlation among the spherical harmonics
coefficients, according to EGM96 [{\it Iorio}, 2002].}; according
to the diagonal part only of the covariance matrix of EIGEN--1S,
it reduces to 15.5$\%$.

It should be noted that, according to [{\it Ries et al.}, 1998],
it would not be entirely correct to automatically extend the
validity of the full covariance matrix of EGM96, which is based on
a multi--year average that spans the 1970, 1980 and early 1990
decades, to any particular time span like that, e.g., of the
LAGEOS--LAGEOS II Lense--Thirring analysis which extends from the
middle to the end of the 1990 decade. Indeed, there would not be
assurance that the errors in the even zonal harmonics of the
geopotential during the time of the LAGEOS--LAGEOS II
Lense--Thirring experiment remained correlated exactly as in the
EGM96 covariance matrix, in view of the various secular, seasonal
and stochastic variations that we know occur in the terrestrial
gravitational field and that have been neglected in the EGM96
solution. Then, the use of the diagonal part only of the
covariance matrix of EGM96 should yield more conservative results.
However, since it turns out that such seasonal effects would
mainly affect just the first even zonal harmonic coefficients of
the geopotential, the uncertainty related to them should be very
small for residual combinations which, by construction, cancel out
just the first even zonal harmonic coefficients of the
geopotential. On the other hand, if we cancel out as many even
zonal harmonics as possible, the uncertainties in the evaluation
of the systematic error based on the remaining correlated even
zonal harmonics of higher degree should be greatly reduced,
irrespectively of the chosen time span. This would have a relevant
importance, e.g., for those even zonal harmonics like $J_6$ and
$J_8$ whose favorable correlation in the covariance matrix of
EGM96 seems to be the source of the perhaps optimistic evaluation
of the systematic error due to the even zonal harmonics of the
geopotential in the case of the LAGEOS--LAGEOS II Lense--Thirring
experiment [{\it Ries et al.}, 1998]. The LAGEOS--LAGEOS II--LARES
proposed combination of [{\it Iorio et al.}, 2002b] would cancel
out, apart from $\delta J_2$ and $\delta J_4$, just $\delta J_6$
and $\delta J_8$.
\section{Conclusions}
The systematic error induced by the mismodelled static part of the
geopotential is the major source of uncertainty in many proposed
or performed tests of General Relativity  in the gravitational
field of the Earth via Satellite Laser Ranging.

In this paper we have explicitly calculated the expressions of the
coefficients of the classical secular precessions of the node and
the perigee induced by the even zonal harmonics of the
geopotential up to degree $l=20$. The explicit expressions of the
classical precessions may be useful, e.g., in getting insights for
designing suitably new relativistic observables which are not too
sensitive to such disturbing effects.

Subsequently, we have compared the mismodelled precessions,
according to EGM96 gravity model and EIGEN--1S preliminary gravity
model which includes 88 days of data from CHAMP, to the general
relativistic gravitomagnetic and gravitoelectric secular trends
affecting the same orbital elements of the existing or proposed
laser--ranged geodetic satellites. Since such satellites are the
natural candidates for a number of relativistic tests in the
gravitational field of the Earth, the presented calculations are
useful in order to get an idea of the level of aliasing induced by
the geopotential on the relativistic signatures of possible
observables which may be built up with the orbital elements of
such satellites. Of course, the obtained results could turn out to
be useful also for other nonrelativistic investigations. While for
the LAGEOS satellites a calculation up to $l=20$ is well adequate,
for the other satellites orbiting at lower altitudes also the
other harmonics of higher degrees should be carefully considered.
The need for reducing the impact of the mismodelled classical
precessions on the relativistic signals is quite apparent.

Finally, we have shown how to calculate explicitly the static
gravitational systematic error on suitably designed combinations
involving the orbital residuals of different satellites.
Preliminary estimates of the errors affecting such combinations
with EGM96 and EIGEN--1S suggest that the future, more accurate
terrestrial global gravity models from CHAMP and GRACE missions
will have a notable impact on the improvement of, among other
things, the precision of many general relativistic tests.
\section*{Acknowledgements}
L. Iorio is grateful to L. Guerriero for his support while at
Bari. Special thanks also to the CHAMP and GRACE team of GFZ at
Potsdam for their kind collaboration.

\end{document}